\documentclass{aa}
\usepackage{times}
\usepackage{graphicx}
\usepackage{xspace}
\usepackage{astron}

\newcommand{\err}[2]{\ensuremath{^{+#1}_{-#2}}\xspace}

\makeatletter
\def\@cite#1#2{#1\if@tempswa , #2\fi}
\def\citey{\def\astroncite##1##2{##1\ (##2)}\@internalcite}
\def\citename{\def\astroncite##1##2{##1}\@internalcite}

\makeatother

\begin{document}

\thesaurus{06(08.16.7 Vela X-1; 08.14.1; 08.13.1; 13.25.5)} 

\title{Vela~X-1 as seen by RXTE}

\author{I.~Kreykenbohm\inst{1} \and P.~Kretschmar\inst{1,3} \and
  J.~Wilms\inst{1} \and R.~Staubert\inst{1} \and E.~Kendziorra\inst{1} \and
  D.E.~Gruber\inst{2} \and W.A.~Heindl\inst{2} \and R.R.~Rothschild\inst{2}}
\offprints{J.~Wilms}
\mail{wilms@astro.uni-tuebingen.de}

\institute{Institut f\"ur Astronomie und Astrophysik -- Astronomie,
  University of T\"ubingen, Waldh\"auser Str. 64, 72076 T\"ubingen, Germany
  \and Center for Astrophysics and Space Sciences, University of
  California, San Diego, La Jolla, CA 92093, USA
  \and INTEGRAL Science Data Center, 1290 Versoix, Switzerland}

\date{Received 14~July~1998 / Accepted 30.~Sept.~1998} 
\maketitle

\begin{abstract}
  We present results from four observations of the accreting X-ray pulsar
  \object{Vela X-1} with the Rossi X-ray Timing Explorer (\textsl{RXTE}) in
  1996 February.
  
  The light curves show strong pulse to pulse variations, while the average
  pulse profiles are quite stable, similar to previous results.  Below
  5\,keV the pulse profiles display a complex, 5-peaked structure with a
  transition to a simple, double peak above $\sim$15\,keV.
  
  We analyze phase-averaged, phase-resolved, and on-pulse minus off-pulse
  spectra. The best spectral fits were obtained using continuum models with
  a smooth high-energy turnover. In contrast, the commonly used power law
  with exponential cutoff introduced artificial features in the fit
  residuals.  Using a power law with a Fermi-Dirac cutoff modified by
  photoelectric absorption and an iron line, the best fit spectra are still
  unacceptable. We interpret large deviations around $\sim$25 and
  $\sim$55\,keV as fundamental and second harmonic cyclotron absorption
  lines. If this result holds true, the ratio of the line energies seems to
  be larger than~2.  Phase resolved spectra show that the cyclotron lines
  are strongest on the main pulse while they are barely visible outside the
  pulses.
  
\end{abstract}

\section{Introduction} 
\begin{figure}
\resizebox{\hsize}{!}{\includegraphics{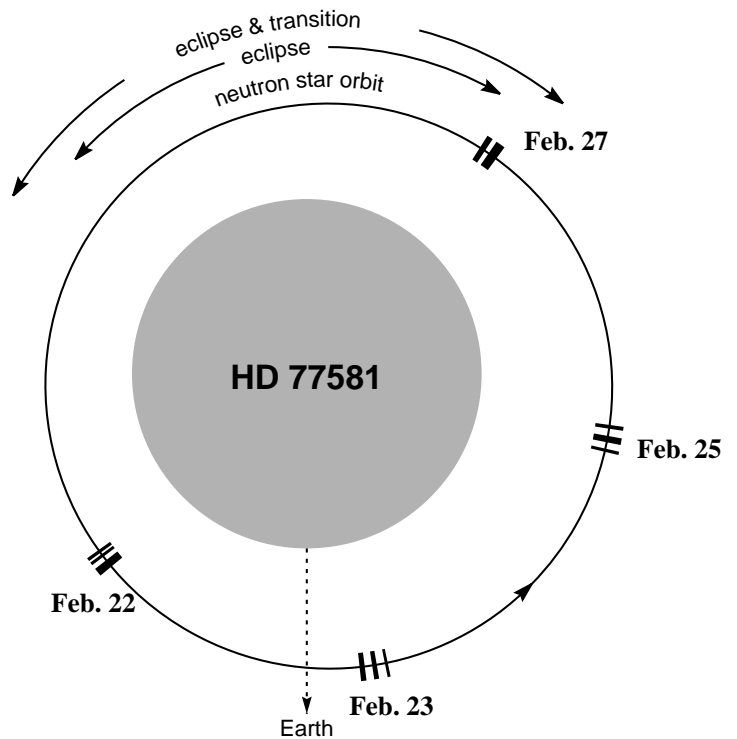}}
\caption{Sketch of the Vela~X-1 system (center of system). The dashes mark the
  approximate orbital positions when the observations were made. The
  movement and the Roche-deformation of HD\,77581 have been neglected.}
\label{fig_velaobs}
\end{figure}

Vela~X-1 (4U\,0900$-$40) is an eclipsing high mass X-ray binary consisting
of the 23\,$M_{\sun}$ B0.5Ib supergiant HD\,77581 and a neutron star with an
orbital period of 8.964\,d (\cite{kerkwijk95a}), corresponding to an
orbital radius of about $1.7\,{\rm R_*}$ (Fig.~\ref{fig_velaobs}). The
system is at a distance of 2.0\,kpc (\cite{sadakane85a}). Due to the
closeness of the neutron star and its companion, the neutron star ($M$
$\sim$1.4\,$M_{\sun}$, \cite{stickland97a}) is deeply embedded in the strong
stellar wind of HD\,77581 ($\dot{M} \sim 4\,10^{-6}\,M_{\sun}/{\rm yr}$,
\cite{nagase86a}). Thus, the system is a prime candidate for the
study of X-ray production via wind accretion.  The X-ray luminosity is
typically $\sim 4\,10^{36}\,{\rm erg/s}$, but can also suddenly change to
less than 10\% of its normal value (\cite{inoue84a}). The reason for
this variability is still unknown. It might be caused by changes in the
accretion rate due to variations in the stellar wind or the formation
of an accretion disk (\cite{inoue84a}). For an in-depth discussion of the
system parameters, see \citey{kerkwijk95a}.

Vela~X-1 is a slow X-ray pulsar with a period of about 283\,seconds
(\cite{rappaport75a}).  Despite significant pulse-to-pulse variations, the
pulse profile averaged over many pulses is quite stable
(\cite{staubert80a}).  The profile changes from a complex five-peaked
structure at energies below $\sim$5\,keV to a simple double peaked
structure at energies above $\sim$15\,keV (\cite{raubenheimer90a}).

The X-ray spectrum of Vela~X-1 is usually described by a power law with an
exponential cutoff at high energies and an iron K$\alpha$ line at 6.4\,keV
(\cite{nagase86a,white83a,ohashi84a}). Below $\sim$3\,keV, a soft excess is
observed (\cite{lewis92a,pan94a}).  The spectrum is further modified by
photoelectric absorption due to a gas stream trailing the neutron star and
due to circumstellar matter (\cite{kaper94a}). This results in
significantly increased absorption at orbital phases $\ga$0.5. In addition,
erratic increases of $n_{\rm H}$ by a factor of 10 are seen at all orbital
phases (\cite{haberl90a}).  At energies above 20\,keV, evidence for
cyclotron absorption features at $\sim$27\,keV and $\sim$54\,keV has been
reported from the High Energy X-Ray Experiment (\textsl{HEXE}) on Mir
(\cite{kendziorra92}), from \textsl{Ginga} (\cite{mihara95a}), and from
\textsl{BeppoSAX} observations (\cite{orlandini98}), but to date no
statistically compelling result on both lines has been obtained.

In this paper we present the results of the analyses of four observations
of Vela~X-1 with \textsl{RXTE}. The observations and data reduction are
described in Sect.~\ref{data}. In Sect.~\ref{lc} we discuss the
light curves and pulse profiles.  Section~\ref{pha} is devoted to the
spectral analysis, including spectral models, phase averaged spectra, phase
resolved spectra, and a comparison with other instruments. Finally, we
discuss the results in Sect.~\ref{discussion}.

\section{Observations and Data Analysis}
\label{data}
We observed Vela~X-1 with \textsl{RXTE} four times, covering nearly a
complete binary orbit from phase 0.3 to 0.9. Each observation was between
5\,ksec and 6\,ksec long (see Table~\ref{tab_velaobs} and
Fig.~\ref{fig_velaobs}). The observations were carried out between 1996
February 22 and 1996 February 27.  The exact time and duration of each
observation are given in Table~\ref{tab_velaobs}. In our analysis we used
data from both \textsl{RXTE} pointing instruments, the Proportional Counter
Array (\textsl{PCA}) and the High Energy X-ray Timing Experiment
(\textsl{HEXTE}).  To avoid scattered photons from the Earth, we used only
data where Vela~X-1 was more than $10^\circ$ above the horizon. To extract
spectra and light curves, we used the standard RXTE analysis software {\sl
  FTOOLS 4.0} and software provided by the \textsl{HEXTE} instrument team
at UCSD.
\begin{table}
\caption{Details of these observations of Vela~X-1 with \textsl{RXTE}. The
  on source time corresponds only to the \textsl{PCA}.}
\label{tab_velaobs}
\begin{tabular}[t]{lllll}
No. & date & JD (center of & orbital & on source \\ & & observation) &
    phase & time \\ \hline 
1 & 1996 February 22 & 2450135.66 & 0.311--0.324 & 6048\,s \\ 
2 & 1996 February 23 & 2450137.30 & 0.492--0.509 & 5696\,s \\ 
3 & 1996 February 25 & 2450139.17 & 0.701--0.717 & 5536\,s \\ 
4 & 1996 February 27 & 2450140.79 & 0.884--0.895 & 5820\,s \\
\end{tabular}
\end{table}

The \textsl{PCA} consists of five co-aligned Xenon proportional counter
units with a total effective area of about 6500\,${\rm cm}^2$ and sensitive
in the energy range from 2\,keV to $\sim$60\,keV (\cite{xte_performance}).
Light curves with 250\,ms resolution and energy spectra in 128 pulse height
channels were collected.  For spectral fitting, we used version 2.2.1 of
the \textsl{PCA} response matrices (Jahoda 1997, priv. comm.).  Due to the
high statistical significance of the data, we applied systematic
uncertainties to the spectra. These were determined from deviations in a
fit to the Crab Nebula and pulsar spectrum and had values of 2.5\% below
5\,keV, 1\% between 5 and 20\,keV, and 2\,\% above 20\,keV.  

Background subtraction in the \textsl{PCA} is usually done using a
background model.  This model does not work yet with this \textsl{RXTE}
epoch~1 data (\cite{stark}), and we were forced to use a different approach
to background subtraction.  Since the \textsl{PCA} is not turned off during
Earth occults an approximation to the average \textsl{PCA} background
spectrum can be obtained by accumulating data measured during the occult.
For this, data where the source was at least 5$^\circ$ below the horizon
was used. Because the background depends on geomagnetic latitude, which
varies throughout the spacecraft orbit, it is obviously not possible to
generate a time variable background estimate from occultation data.
Therefore, the light curves presented here are not background subtracted.
To check the validity of our approach in the background
subtraction we compared spectra obtained during Earth occults from later
post gain-shift data with background spectra estimated for these data. In
general, the agreement between the estimated and the measured spectrum was
good for these data. This suggests that our approach will result in an
adequate background subtraction, especially considering the fact that our
data are completely source dominated.  Finally, due to uncertainties in the
response matrix above 30\,keV and since the high energy channels might be
background dominated, we limited the PCA energy range from 3 to 30\,keV.

The \textsl{HEXTE} consists of two clusters of four NaI/CsI-phoswich
scintillation counters, sensitive from 15 to 250\,keV.  A full description
of the instrument is given by \citey{rothschild97a}.  Our observations were
made before the loss of energy information from one detector, so we have
the full \textsl{HEXTE} area. 
Background subtraction is done by
source-background rocking of the two clusters which provides a direct
measurement of the \textsl{HEXTE} background during the observation such
that no background model is required. This method has measured systematic
uncertainties of $<1\%$ (\cite{rothschild97a}).  
We used the standard
response matrices dated 1997~March~20 and treated each cluster individually
in the data analysis.
We ignored channels below 20, and
to improve the statistics of individual energy bins, we rebinned the raw
($\sim$1\,keV wide) energy channels as enumerated in Table~\ref{tab_rebin}.
\begin{table}
\caption{Rebinning of the \textsl{HEXTE}-data. The factor is the number of
instrument channels rebinned into a new channel.}
\label{tab_rebin}
\begin{tabular}{cr@{ \ \ -- }rr} Group & low & high & factor \\
\hline 1 & 20 & 39 & 4 \\ 2 & 40 & 63 & 6 \\ 3 & 64 & 103 & 8 \\ 4 & 104 &
153 & 10 \\ 5 & 154 & 255 & 20 \\
\end{tabular}
\end{table}

\subsection{Spectral Models}
\label{models}
\subsubsection{Continuum Models}
\label{s_cont}
Despite two decades of work, there still exist no convincing theoretical
models for the shape of the X-ray spectrum in accreting X-ray pulsars (cf.\ 
\cite{harding94a}, and references therein).  Therefore, we are forced to use
empirical spectra in the modeling process.

As in previous observations with \textsl{Ginga} and \textsl{HEXE}
(\cite{mihara95a,kretschmar97c}), we found it impossible to fit the spectra
with thermal bremsstrahlung or blackbody spectra, while we were able to fit
our spectra with a power law modified by a high energy cutoff.  The
standard version of this cutoff (\textsl{PLCUT} henceforth) is analytically
realized as follows (\cite{white83a}):
\begin{equation} 
C(E) = E^{-\alpha} \times \left\{
\begin{array}{l@{\mbox{for }}l} 
  1 & E < E_{\rm cut} \\ \exp{-\left(\frac{E-E_{\rm cut}}{E_{\rm fold}}\right)}
& E > E_{\rm cut}
\end{array}\right.\label{eq_plcut}
\end{equation}
In this model, the high energy cutoff is switched on at the cutoff energy,
$E_{\rm cut}$, and the derivative is discontinuous at this point.  The
source spectrum, however, most probably does not have such a sudden
turnover. Instead, the importance of the cutoff might increase steadily
from insignificance at low energies until it dominates the power law above
the cutoff energy. If such a spectrum is fit with the \textsl{PLCUT} model,
the resulting residuals are likely to mimic a line feature at $E_{\rm
  cut}$, and might lead to erroneous conclusions (\cite{kretschmar96b}).
 
\begin{figure}
\resizebox{\hsize}{!}{\includegraphics{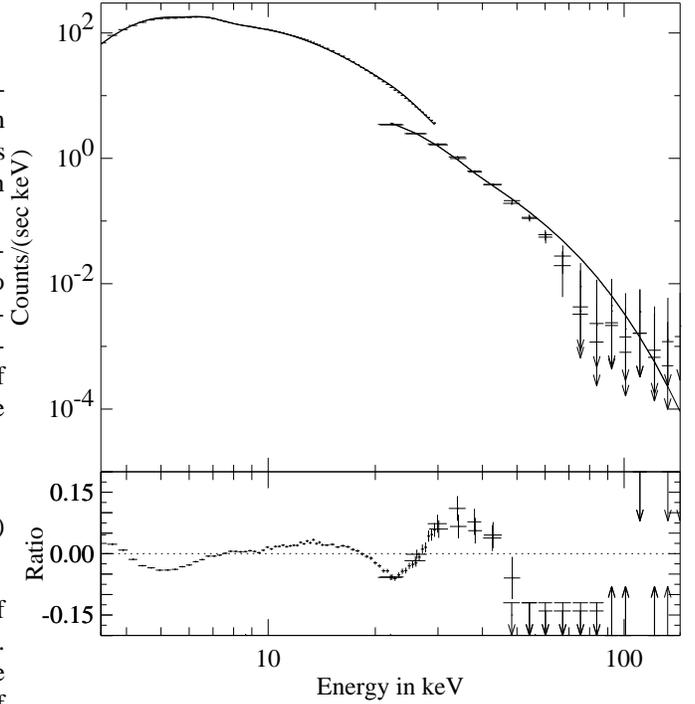}}
\caption{A simulated spectrum based on a power law with a smooth cutoff. This
spectrum has been fit with the \textsl{PLCUT} model with
$E_{\rm cut}\approx$20\,keV. The ratio is given by (data-model)/data.}
\label{fig_simul}
\end{figure}

To demonstrate the effects of applying the \textsl{PLCUT} model to spectra
with a smooth continuum, we simulated a \textsl{PCA} spectrum using a power
law with a \textit{smooth} exponential turnover, including an iron line at
6.4\,keV and photoelectric absorption. The spectral parameters for the
simulation were chosen to be appropriate for Vela~X-1.  We then tried to
use the standard \textsl{PLCUT} model to describe this simulated spectrum.
The result is shown in Fig. \ref{fig_simul}. The fit features a deep
absorption line-like feature at about 20\,keV and some sort of additional
cutoff (or a second absorption line) above 40\,keV. Note that $E_{\rm cut}$
has almost the same value as the energy of the first line-like feature and
both are strongly correlated.  These two line-like features might easily be
misinterpreted as a cyclotron absorption line and its second harmonic.

It is obvious, therefore, that the explicit form of the continuum is
crucial for the detection of absorption (or emission) lines like those
resulting from interactions between Larmor electrons and X-rays at the
cyclotron energy.  But as the real continuum is not known the only thing we
can do is to avoid those continua from which we know that they introduce
artificial features in the fit residuals.

We used the so-called Fermi-Dirac cutoff (\textsl{FDCO} in the following) first
mentioned by \citey{tanaka86a}, which has a smooth and continuous turnover.
\begin{equation}\label{eq:fdco} C(E) = E^{-\alpha} \times
  \frac{1}{\exp{\left(\frac{E-E_{\rm cut}}{E_{\rm fold}} 
\right)}+1}\label{eq_fdc}\end{equation}
Due to historical reasons, the parameters of the \textsl{PLCUT}- and the
\textsl{FDCO}-model have the same names, but we stress that it is not
possible to compare them as the continua are different.

Another spectral model avoiding a sharp turnover is the 
Negative Positive Exponential (\textsl{NPEX}) model introduced by
\citey{mihara95a}: 
\begin{equation} C(E) = (A_1\cdot E^{-\alpha_1} + A_2\cdot E^{+\alpha_2})\times
  \exp\left(-\frac{E}{E_{\rm fold}}\right)\label{eq_npex}\end{equation}
\noindent where $\alpha_1$, $\alpha_2 > 0$. This model can fit very complex
continuum forms, including dips, nearly flat areas, and bumps strongly
depending on the normalization of the power law with the positive photon
index.
If the second photon index is fixed at a value of $\alpha_2=2.0$ this
model is a simple analytical approximation of a Sunyaev-Titarchuk
Comptonization spectrum (\cite{sunyaev80a}). \citey{mihara95a} found that
this restricted form of the model was able to describe a large variety
of X-ray pulsar spectra.

We fit our spectra with both the \textsl{FDCO} and the \textsl{NPEX} models
and found that both provided reasonable descriptions of the \emph{continua}
(see Sect.~\ref{ginga}). 
In comparison, the \textsl{FDCO} spectrum appeared to give more reliable
results. With the \textsl{NPEX} model the normalization of the second power
law was often set to practically zero by the fit software, resulting in a
standard power law with exponential cutoff. We therefore decided to use
the \textsl{FDCO} model for detailed fits.

\subsubsection{Absorption Line Models}
Two models are generally employed to describe cyclotron absorption line
profiles.  In both cases, the absorption is described by an exponential
term of the form $\rm e^{-\tau(E)}$, where $\tau$ is the optical depth of
the absorption as a function of energy.  This term is multiplied by a
continuum model (Sect.~\ref{s_cont}). The first line model is a simple
Gaussian profile for the optical depth.  The second has a Lorentzian
profile of the form (\cite{herold79a})
\begin{equation} 
\tau(E) = \frac{A({WE}/E_{\rm cyc})^2}{(E-E_{\rm cyc})^2 + W^2}
\end{equation}  
In this case, $W$ is the line width, $A$ the depth, and $E_{\rm cyc}$ the
line energy. The fundamental line and its harmonics may be included by
employing multiple absorption terms, with or without constraining the
relative values of the parameters (e.g. the relative line energies or
depths).  The statistics of our observations are insufficient to choose
between these two models, and all fits reported here employ the Lorentzian
profile.  The quoted widths thus refer to the parameter $W$.

\begin{figure}
\resizebox{\hsize}{!}{\includegraphics{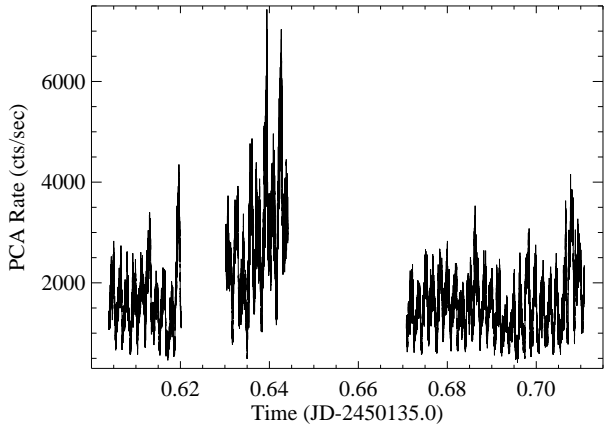}}
\caption{\textsl{PCA} light curve of observation~1. The first gap is due to the
  South Atlantic Anomaly and the second to an Earth occultation.}
\label{fig_flare}
\end{figure}

\begin{figure}
\resizebox{\hsize}{!}{\includegraphics{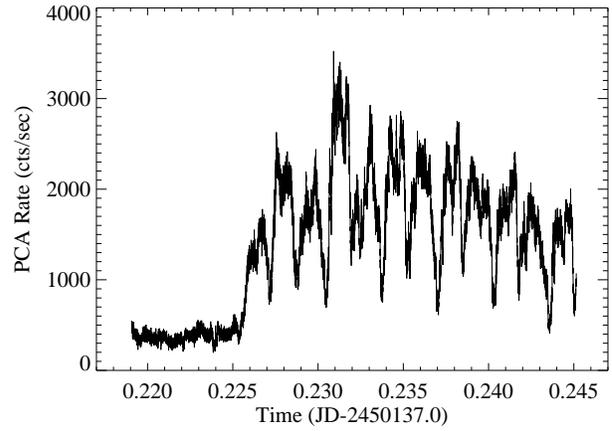}}
\caption{The beginning of the \textsl{PCA} light curve of observation~2 with an
  extended period of time where no pulsations are observed.}
\label{obs2off}
\end{figure}

\section{Light curves and Pulse Profiles}
\label{lc}
The average PCA count rate in our four observations is between 1800\,cts/s in
observation~1 and 1000\,cts/s in observations~3, corresponding to a normal
to low flux level. In addition to this normal behavior we observed two special
events: 

1.~An increase of the count rate in a flare-like event with a maximum
\textsl{PCA} count rate of about 7000~cts/s during observation~1 (see Fig.
\ref{fig_flare}). Due to an Earth occultation, we did not observe the end
of the flare: after the occultation, Vela~X-1 was in its normal state
again.

2.~An interval of about 550\,s duration at the beginning of observation~2,
where the average total \textsl{PCA} count rate (i.e., including the
background contribution of $\sim$150\,cts/s) was below 400~cts/s
(Fig.~\ref{obs2off}).  During this interval, no pulsations were observed.
After that, the count rate increased and reached its normal value of about
1800\,cts/s within one pulse period. This behavior has been observed
before by \citey{inoue84a} and \citey{lapshov92a}, but its cause is still
unknown and further investigations are needed.

\begin{figure}
\resizebox{\hsize}{!}{\includegraphics{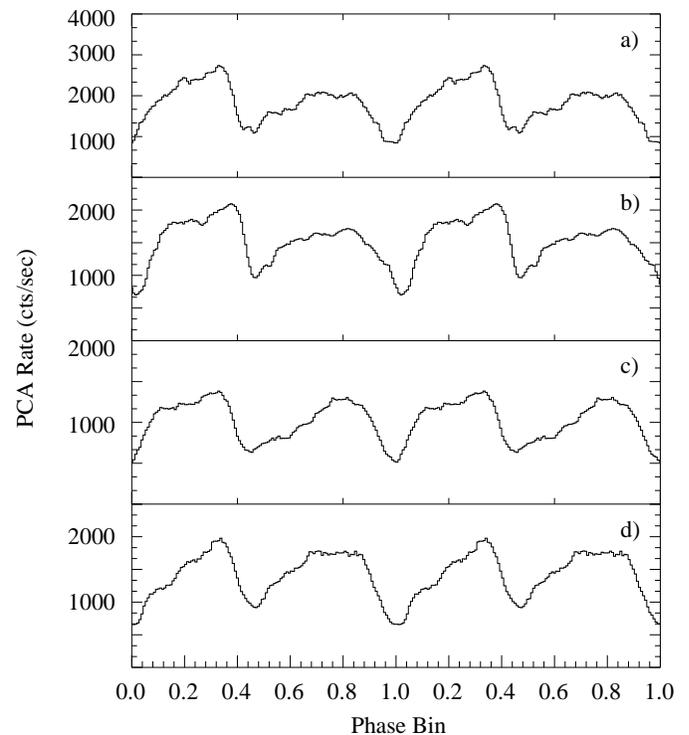}}
\caption{Energy averaged \textsl{PCA} pulse profiles for observations 1 to 4
(panels {\bf a} to {\bf d}). The profiles were derived by folding
with a period of 283.0\,sec in 128 phase bins over the entire energy range.}
\label{fig_avgprofil}
\end{figure}

\begin{figure}
\resizebox{\hsize}{!}{\includegraphics{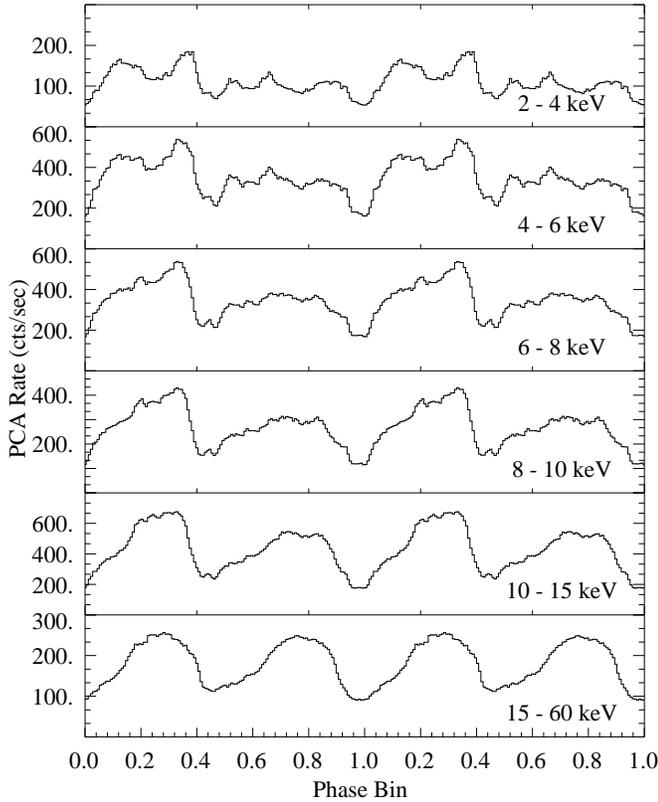}}
\caption{Observation~1 pulse profiles in 6 energy bands. The profiles
  consist of 128 phase bins.} 
\label{fig_enprofile2}
\end{figure}

We derive an average pulse period of $\rm 283.4 \pm 1.7$\,s for all four
observations, consistent with the value reported by \textsl{BATSE} for this
epoch (\cite{batse96}).  In Fig.~\ref{fig_avgprofil}, pulse profiles for
the whole \textsl{PCA} energy range are shown.

Despite the fact that these pulse profiles were obtained at different
orbital phases, they are quite similar, with differences possibly somewhat
affected by the low number of pulses added (about 20).  Energy resolved
pulse profiles (Fig.~\ref{fig_enprofile2}) illustrate the complex behavior
found by \citey{raubenheimer90a}: At energies below $\sim$5\,keV, a very
complex five peak structure is present. The main pulse consists of two
peaks with a deep gap between them, while the secondary pulse consists of
three peaks.  At higher energies this structure simplifies until the pulse
profile consists of two pulses with nearly the same intensity above
$\sim$15\,keV.

\section{Spectral Results} \label{pha} \subsection{Phase Averaged Spectra}
In addition to the \textsl{FDCO} model we allowed for photoelectric
absorption and an iron fluorescence line at 6.4\,keV in our spectral fits.
The iron line is thought to originate in cold, circumstellar material where
some X-rays are reprocessed.  While we found a relatively narrow line in
observations~1 and~4 it was impossible to obtain an acceptable fit with a
narrow line for observations~2 and~3. Instead we obtained line widths of
about 1.3\,keV (obs.~2) to 1.6\,keV (obs.~3;
cf.~Table~\ref{tab_avgresults}). Similar values have been found in some
\textsl{EXOSAT} observations (\cite{gottwald95a}),
later ASCA measurements did not find evidence for such small line
widths. We caution, however, that our broad line widths might also be due
to the finite energy resolution of the detector in combination with
response matrix features.
 Due to the gas stream
trailing the neutron star, the photoelectric absorption increased
dramatically between the second and the third observations.  Since the
lower energy limit of our spectra is 3\,keV we are not able to observe the
soft-excess below 4\,keV reported by \citey{lewis92a} and
\citey{haberl94a}.

\begin{figure}
\resizebox{\hsize}{!}{\includegraphics{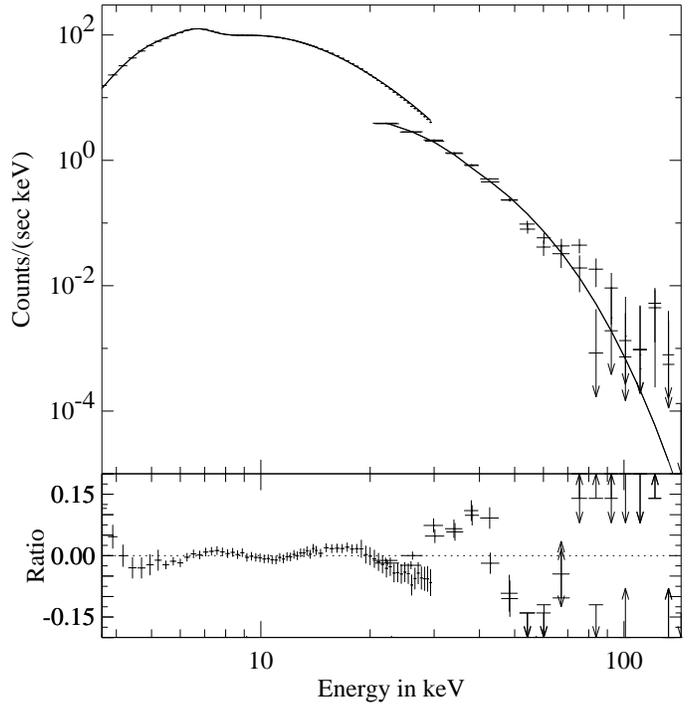}}
\caption{Fit with the \textsl{FDCO} model, photoelectric absorption, and an
  additive iron line at 6.4\,keV to the observation~4. The upper panel
  shows the data and the model for the \textsl{PCA} and for both
  \textsl{HEXTE}-clusters. The lower panel shows the corresponding ratios.}
\label{fig_obs5_nocyc}
\end{figure}

Still, this description of the data is not acceptable, showing significant
deviations near 25\,keV and 55\,keV.  We interpret these features as
fundamental and second harmonic cyclotron absorption lines. Therefore we
included a cyclotron absorption component in our fits.

Initially, we coupled the fundamental and second harmonic line energies and
widths by a factor of 2. Since the statistics of the data were inadequate
to constrain both the widths and depths of both lines, we also fixed the
width of the fundamental line to a value of 5\,keV (\cite{harding91a}).
While formally acceptable fits for all observations were found, the second
harmonic line had a depth of nearly zero.  This contradicts the apparent
presence of a feature near 55~keV seen in the residuals of most fits (see
Fig.~\ref{fig_obs5_nocyc}). We concluded that the second harmonic line is
coupled to the fundamental line with a factor \textit{greater} than 2.0.

Therefore we introduced a variable coupling factor as a new free parameter.
We used the same parameter to specify the ratios of line energies and
widths (Harding 1997, priv. comm.).  Using this modified model, we found
the fundamental line in all four observations, however with variable depth
(see Table~\ref{tab_avgresults}). The energy of the fundamental cyclotron
line is found between 22\,keV and 25\,keV. A second harmonic line is seen
in observations~1 and~4; however, a second line did not improve the fit for
observations~2 and~3. Where a line was present, it was coupled to the
fundamental line by a factor of 2.3 to 2.4 (corresponding to a line energy
of $\sim 56$\,keV) instead of 2.0.  Table~\ref{tab_avgresults} summarizes
the results of the fits to all four observations. To test the significance
of both lines we used the F-test (\cite{bevington69}).  The results are
given in Table~\ref{tab_avgftest}.  While the fundamental line is very
significant in all four observations, this is not the case for the harmonic
line.  Only for observation~4 is it significant. Note that all
uncertainties quoted in this paper represent 90\% confidence level for a
single parameter.

\begin{figure}
\resizebox{\hsize}{!}{\includegraphics{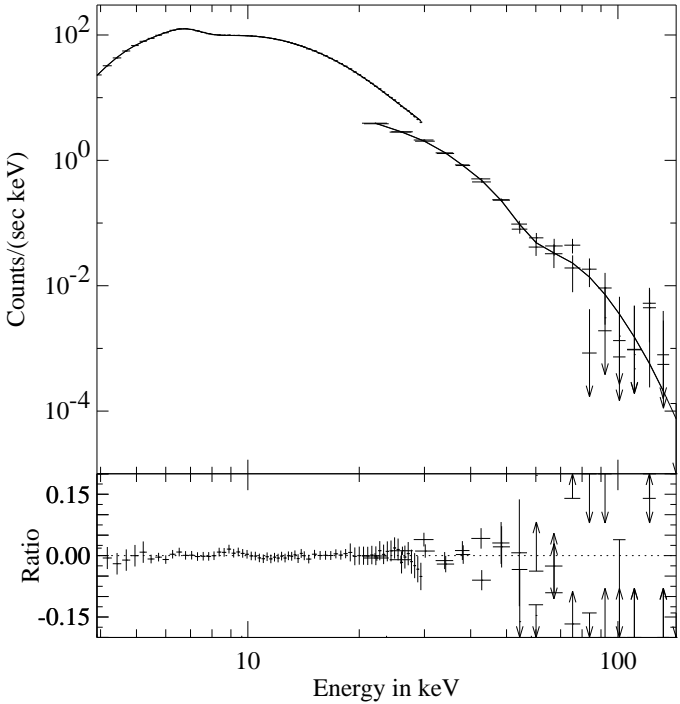}}
\caption{Same as Fig.~\ref{fig_obs5_nocyc}, but with two cyclotron
  absorption lines.}
\label{fig_obs5_2cyc}
\end{figure}

\begin{table}
\caption{Parameters of the \textsl{FDCO} model with an Iron line,
  photoelectric absorption, and cyclotron absorption lines. The symbols
  used are: 
  $N_{\rm H}$: neutral Hydrogen column, in $10^{22}\,{\rm cm}^{-2}$,
  $E_{\rm cut}$, $E_{\rm fold}$: parameters of the FDCO model
  (cf.~eq.~\ref{eq:fdco}), Fe-$\sigma$: width of the Gaussian iron line in
  keV (the energy of the line was fixed at 6.4\,keV), $E_{\rm cyc}$,
  Depth~1, and Depth~2: Parameters of the cyclotron line, the width of
  which was fixed to 5\,keV, Factor: Coupling factor between the
  fundamental cyclotron line energy and the first harmonic, DOF:
  Degrees of Freedom of the fit. } 
\label{tab_avgresults}
\begin{tabular}
{lr@{\hspace*{0.3em}}lr@{\hspace*{0.3em}}lr@{\hspace*{0.3em}}lr@{\hspace*{0.3em}}l}
& \multicolumn{2}{c}{Obs.~1}
&\multicolumn{2}{c}{Obs.~2} & \multicolumn{2}{c}{Obs.~3}
&\multicolumn{2}{c}{Obs.~4} \\ 
\hline 
\rule{0pt}{2.5ex} $N_{\rm H}$ & 9.2 &\err{0.5}{0.7} & 4.2 & \err{0.6}{0.6} &
22.3 &
\err{0.6}{0.5} & 30.1 & 
\err{1.1}{1.0} \\ 
\rule{0pt}{3ex} $\alpha$ & 1.29&\err{0.04}{0.07}&1.67&\err{0.03}{0.04}&
1.67&\err{0.05}{0.03}& 0.94&\err{0.04}{0.04}\\ 
\rule{0pt}{3ex} $E_{\rm cut}$ & 33.4&\err{1.4}{0.5}& 39.9&\err{1.1}{1.2}&
37.7&\err{1.4}{2.0}& 21.0&\err{5.1}{7.0} \\
 \rule{0pt}{3ex} $E_{\rm fold}$ & 8.3&\err{1.1}{1.4} & 6.9 &\err{0.8}{0.8} &
 9.1&\err{1.1}{0.8} & 13.3&\err{0.4}{0.5}\\  
Fe-$\sigma$ & 0.61 &\err{0.10}{0.08}& 1.29&\err{0.14}{0.14}&
1.62&\err{0.17}{0.32}& 0.60 &\err{0.10}{0.10}\\ 
\rule{0pt}{3ex} $E_{\rm cyc}$ & 24.1&\err{0.7}{0.5} & 22.6&\err{1.6}{1.5} &
21.6&\err{0.6}{0.9} & 23.3&\err{0.4}{0.3} \\ 
\rule{0pt}{3ex} Depth 1 & 0.18&\err{0.05}{0.01} & 0.07&\err{0.03}{0.03} &
0.16&\err{0.02}{0.03} & 0.14&\err{0.02}{0.02} \\ 
\rule{0pt}{3ex} Depth 2 & 0.32&\err{0.10}{0.79} & \multicolumn{2}{c}{---} &
\multicolumn{2}{c}{---} & 1.37 &\err{0.28}{0.24} \\ 
\rule{0pt}{4ex} Factor & 2.33&\err{\infty}{0.22} & \multicolumn{2}{c}{---}
& \multicolumn{2}{c}{---} & 2.38 &\err{0.10}{0.10} \\ 
\rule[-1.5ex]{0pt}{3ex} Constant & 0.78 & \err{0.00}{0.01} & 0.71 &
\err{0.01}{0.01} & 0.72 & \err{0.01}{0.01}& 0.78& \err{0.01}{0.01} \\ 
\hline 
\rule[-1.0ex]{0pt}{3ex} $\chi^2$ (DOF) & \multicolumn{2}{c}{66 (94)} &
\multicolumn{2}{c}{55 (96)} & \multicolumn{2}{c}{68 (95)} &
\multicolumn{2}{c}{63 (92)} \\ 
\end{tabular}
\end{table}

\begin{table}
\caption{Comparison of $\chi^2$ and F-test values of all four observations
for fits without, with one, and two cyclotron absorption lines.
$Q$(F) is the probability that the improvement is accidental.}
\label{tab_avgftest}
\begin{tabular}{llllll} 
& Obs.~1 & Obs.~2 & Obs.~3 & Obs.~4 \\ 
\hline 
no lines& 221.2 (97) & 69.4 (98) & 128.4 (98) & 279.4 (96)\\ 
1 line& 67.1 (95) & 55.3 (96) & 68.5 (96) & 104.7 (94) \\ 
\hfill $\Longrightarrow Q$(F) &0.0\,\% & 0.0018\,\% &$<10^{-11}$\,\% &
0.0\,\% \\ 
2 lines & 65.6 (93) & --- & --- & 63.4 (92) \\ 
\hfill $\Longrightarrow Q$(F) &34\,\% & --- & --- & $<10^{-8}$\,\% \\
\end{tabular}
\end{table}

\subsection{Phase resolved Spectra}
\subsubsection{Fine Phase Bins}
\label{ten}

\begin{figure}
\resizebox{\hsize}{!}{\includegraphics{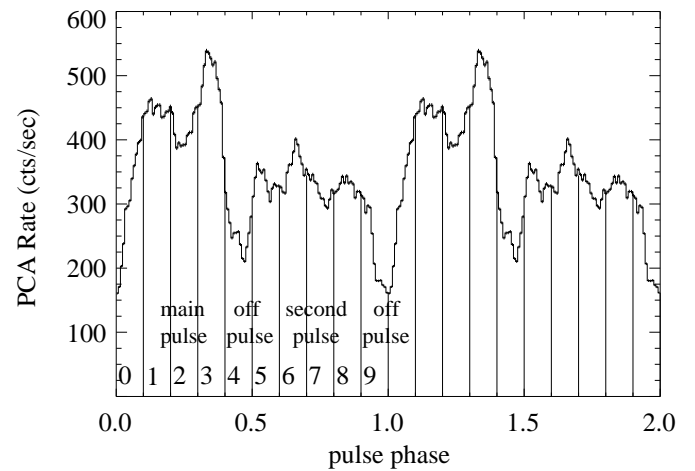}}
\caption{Definition of the ten phase bins for observation~1.} 
\label{fig_phasen}
\end{figure}

In order to study spectral variations with the pulse phase, we split the
pulse in 10 phase bins as shown in Fig.~\ref{fig_phasen}.

With the reduced statistics of these phase resolved spectra, we were unable
to constrain the presence of a second harmonic line.  We therefore restrict
this analysis to the continuum and fundamental line.
\begin{figure}
\resizebox{\hsize}{!}{\includegraphics{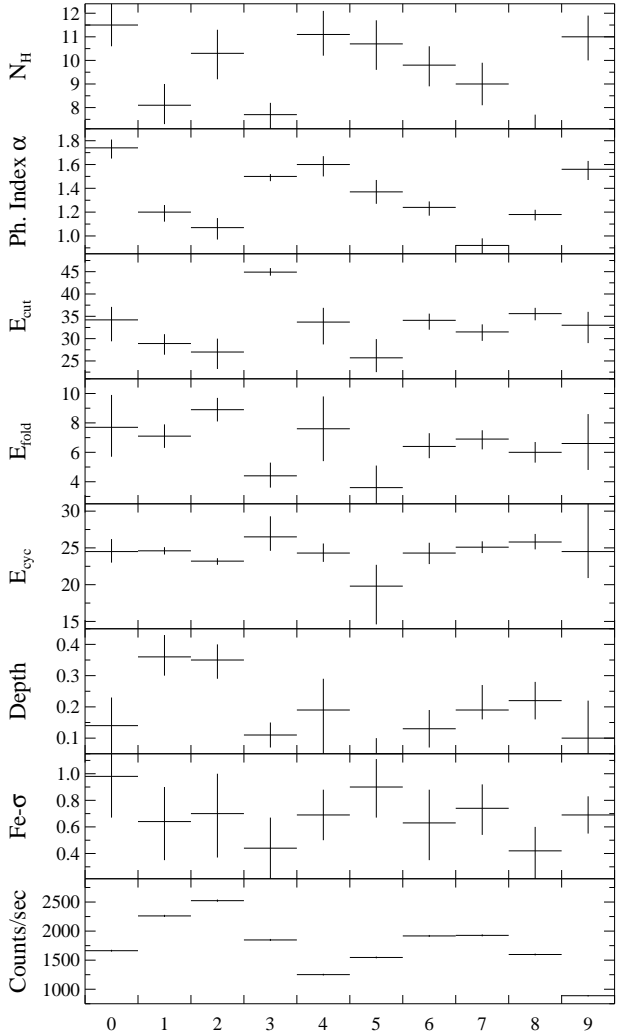}}
\caption{Variation of the spectral parameters over the pulse of observation
  number~1. The vertical bars indicate the uncertainty at the 90\% level.}
\label{fig_resvar}
\end{figure}

The behavior with phase is the same for all four observations.  We
therefore discuss only observation~1 (see Fig.~\ref{fig_resvar}).  The
cutoff energy $E_{\rm cut}$, folding energy $E_{\rm fold}$, cyclotron
absorption energy $E_{\rm cyc}$, and the width of the iron line show no
significant variation over the pulse phase. In contrary, the photon index
and the depth of the cyclotron line are strongly correlated with the pulse
phase.  The variation of the photon index (Fig.~\ref{fig_resvar}) describes
a spectral hardening during the pulse peaks. This hardening is especially
significant in the second pulse.  The depth of the fundamental line varies
also significantly over the pulse phase. The cyclotron absorption is
strongest in the pulse peaks, especially the main pulse, while it is quite
weak or absent outside the pulses.

\subsubsection{Main Pulse, Secondary Pulse, and Off-Pulse}
\label{sec_main}
To study the second harmonic cyclotron absorption line, it is necessary to
increase the statistical significance of the bins above $\sim$50\,keV.
Therefore, we combined the spectra of the ten phase bins into spectra of
the main pulse, the secondary pulse, and the off-pulse according to
Fig.~\ref{fig_phasen}. The parameters resulting from fits to these spectra
are shown in Table~\ref{tab_hno}.  The fundamental cyclotron absorption line
is deepest in the main pulse while it is relatively shallow in the
secondary pulse and barely detectable in the off-pulse.

\begin{table*}
\caption{The fit-results of the fits to spectra of the main, the secondary,
and the off-pulse. The width of the fundamental cyclotron line
was fixed to 5\,keV and the energy of the iron line to 6.4\,keV. The
symbols have the same meaning as in Table~\ref{tab_avgresults}.}
\label{tab_hno}
\begin{tabular}{l@{\hspace*{-3.5ex}}cr@{\hspace*{0.3em}}lr@{\hspace*{0.3em}}lr@{\hspace*{0.3em}}lr@{\hspace*{0.3em}}lr@{\hspace*{0.3em}}lr@{\hspace*{0.3em}}l}
\multicolumn{2}{c}{ } & \multicolumn{6}{c}{Observation~1} &
\multicolumn{6}{c}{Observation~2} \\ 
\hline 
Parameter && \multicolumn{2}{c}{Main Pulse} & \multicolumn{2}{c}{Sec. Pulse} &
\multicolumn{2}{c}{Off-Pulse} &
\multicolumn{2}{c}{Main Pulse} & \multicolumn{2}{c}{Sec. Pulse} &
\multicolumn{2}{c}{Off-Pulse} \\ 
\hline $N_{\rm H}$ &[ $10^{22}\,{\rm cm}^{-2}$]&8.5&\err{0.8}{0.9} &
8.1&\err{0.8}{0.8} &
11.4&\err{0.9}{0.9} &4.5&\err{0.7}{0.6} & 3.8&\err{0.4}{0.5} & 3.4
&\err{0.5}{0.4}\\ 
$\alpha$ && 1.20&\err{0.07}{0.08} & 1.07&\err{0.05}{0.05} &
1.61&\err{0.06}{0.09} & 1.59&\err{0.05}{0.05} & 1.55&\err{0.02}{0.03} &
1.81&\err{0.02}{0.02} \\ 
$E_{\rm cut}$ & [ keV] & 32.4&\err{3.3}{4.6} & 33.6&\err{2.9}{5.0} &
32.4&\err{2.4}{3.3} & 35.1&\err{2.0}{2.6} & 40.7&\err{0.8}{1.0}
&44.9&\err{1.4}{1.2} \\ 
$E_{\rm fold}$ & [ keV] &11.3&\err{1.6}{1.5} & 7.6&\err{1.4}{1.3} &
8.0&\err{1.3}{1.2} & 7.4&\err{0.9}{0.8} &6.0&\err{0.8}{0.7} &
4.6&\err{1.1}{1.0} \\ 
Depth 1 && 0.28&\err{0.04}{0.03}& 0.18&\err{0.04}{0.04} & 0.10 &
\err{0.06}{0.06} & 0.31&\err{0.06}{0.06} &\multicolumn{2}{c}{---} &
\multicolumn{2}{c}{---} \\ $E_{\rm cyc}$ & [
keV] &24.3&\err{0.4}{0.4} &24.9&\err{1.0}{1.0} & 23.4&\err{1.5}{1.7} &
23.3&\err{0.6}{0.5} & \multicolumn{2}{c}{---}& \multicolumn{2}{c}{---} \\ 
Depth 2 &&1.55&\err{0.72}{0.64} & \multicolumn{2}{c}{---} &
\multicolumn{2}{c}{---} &\multicolumn{2}{c}{---} &
\multicolumn{2}{c}{---} & \multicolumn{2}{c}{---} \\ 
Factor && 2.40&\err{0.21}{0.13} &\multicolumn{2}{c}{---} &
\multicolumn{2}{c}{---} & \multicolumn{2}{c}{---} &
\multicolumn{2}{c}{---} & \multicolumn{2}{c}{---} \\ 
Fe--$\sigma$ & [keV] &0.54&\err{0.20}{0.23} & 0.58&\err{0.14}{0.15}
&0.79&\err{0.16}{0.17} & 1.42&\err{0.15}{0.15} & 1.31&\err{0.12}{0.12}
&1.03&\err{0.17}{0.16} \\ 
Const. && 0.78&\err{0.01}{0.02} &0.79&\err{0.01}{0.02} & 0.77
&\err{0.02}{0.01} &0.72&\err{0.02}{0.02}&0.73&\err{0.01}{0.02} &
0.68&\err{0.02}{0.01} \\ 
\hline 
$\chi^2$ &(DOF) &63.9 &(94) &76.9 &(96) & 50.3 &(96)& 59.0&(96) & 87.6
&(98) & 78.9 &(98) \\ 
\multicolumn{14}{c}{\rule{0pt}{0.03\textheight}}\\ 
\multicolumn{2}{c}{ } &\multicolumn{6}{c}{Observation~3} &
\multicolumn{6}{c}{Observation~4}\\ 
\hline 
Parameter && \multicolumn{2}{c}{Main Pulse} & \multicolumn{2}{c}{Sec. Pulse} &
\multicolumn{2}{c}{Off-Pulse}& \multicolumn{2}{c}{Main
  Pulse} & \multicolumn{2}{c}{Sec. Pulse} & \multicolumn{2}{c}{Off-Pulse} \\ 
\hline 
$N_{\rm H}$ & [$10^{22}\,{\rm cm}^{-2}$]&22.5&\err{1.6}{1.2} &
21.3&\err{1.1}{1.0} & 21.6
&\err{0.9}{0.9} &35.8 &\err{1.0}{1.1} & 29.2&\err{1.2}{1.0} &
30.5&\err{1.5}{1.7} \\ 
$\alpha$ && 1.57&\err{0.11}{0.15} &1.43&\err{0.08}{0.09} &
1.85&\err{0.06}{0.06} & 1.28&\err{0.03}{0.07} &0.76&\err{0.04}{0.05} &
1.27&\err{0.10}{0.13} \\ 
$E_{\rm cut}$ & [ keV]&25.9&\err{7.0}{14.0} & 32.9&\err{3.7}{5.3} &
44.7&\err{1.9}{2.4} &31.5&\err{1.7}{0.8} & 27.2&\err{1.7}{0.8} &
29.7&\err{5.9}{9.2} \\ 
$E_{\rm fold}$ &[ keV] &12.2&\err{2.6}{2.3} & 10.1&\err{1.8}{1.6} &
6.4&\err{2.6}{2.0} &11.2&\err{0.7}{0.6} & 9.6&\err{1.0}{0.9} &
13.9&\err{2.0}{2.0} \\ 
Depth 1&& 0.20&\err{0.06}{0.06} & 0.10&\err{0.04}{0.04} & 0.11
&\err{0.04}{0.03} &0.29&\err{0.03}{0.04} & 0.16&\err{0.03}{0.03} &
0.10&\err{0.03}{0.04} \\ 
$E_{\rm cyc}$ & [ keV] &21.5&\err{0.7}{0.7} & 20.9&\err{1.6}{1.6}
&19.9&\err{1.5}{1.4} & 23.7&\err{0.4}{0.3}& 24.2&\err{1.0}{0.9}
&23.4&\err{0.7}{0.6} \\ 
Depth 2 &&\multicolumn{2}{c}{---} & \multicolumn{2}{c}{---} &
\multicolumn{2}{c}{---}&1.41&\err{0.67}{0.61} &
0.76&\err{0.34}{0.35} & 1.49& \err{0.75}{0.36} \\ 
Factor && \multicolumn{2}{c}{---} & \multicolumn{2}{c}{---} &
\multicolumn{2}{c}{---} & 2.52&\err{0.15}{0.15}
&2.12&\err{0.17}{0.11} & 2.35&\err{0.20}{0.12} \\ 
Fe--$\sigma$ & [ keV] &1.51 & \err{0.23}{0.23} & 1.61&\err{0.20}{0.19} &
1.49 &\err{0.17}{0.17} &0.80 &\err{0.50}{0.48} &0.46 &\err{0.16}{0.16} &
0.59& \err{0.11}{0.12} \\ 
Const && 0.74&\err{0.02}{0.02}& 0.72&\err{0.02}{0.01} &
0.70&\err{0.02}{0.03}& 0.77&\err{0.01}{0.02}&0.79&\err{0.01}{0.02} &
0.78&\err{0.01}{0.01} \\ 
\hline 
$\chi^2$ & (DOF)& 79.8 &(96) & 43.3 &(96) &63.5 &(96) & 76 &(94) & 60.5
&(94) & 57.9 &(94) \\ 
\end{tabular}
\end{table*}

\begin{figure}
\resizebox{\hsize}{!}{\includegraphics{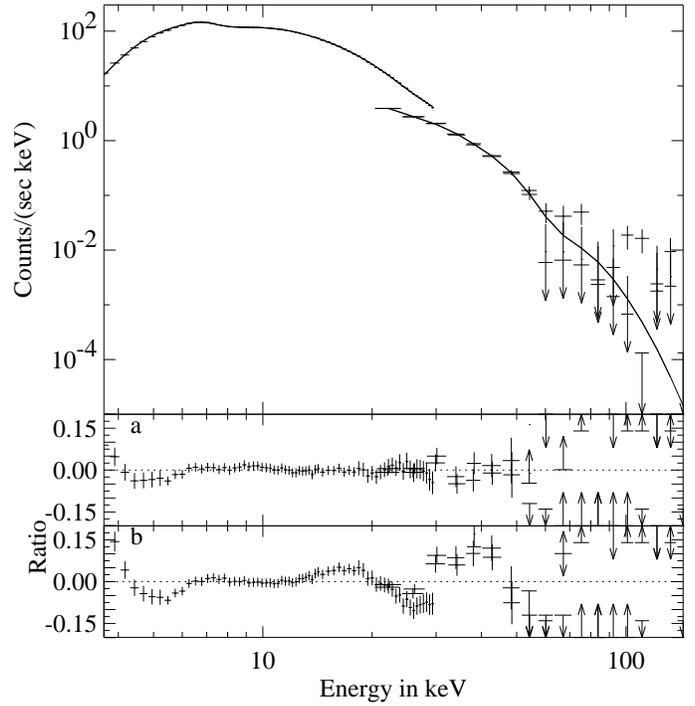}}
\caption{{\bf a and b} Fit to the spectrum of the main pulse of
  observation~4. The upper panel shows the folded spectrum and model with
  both cyclotron lines. The two lower panels show the residuals for the
  cases {\bf a} with both and {\bf b} without cyclotron absorption lines.}
\label{fig_mpuls}
\end{figure}

For the second harmonic line, the results are less straightforward. It is
only found in the main pulse spectrum of observations~1 and~4. In both
cases it is deep and significant at the 99.9\,\% level. A fit to the
spectrum of the main pulse of observation~4 and the corresponding residuals
for models both with and without cyclotron absorption lines is shown in
Fig.~\ref{fig_mpuls}.  In the case of the other two observations the fit
does not improve significantly with the inclusion of the harmonic line.
This is almost certainly due to the low statistical quality of the data
above $\sim$60\,keV.  The harmonic line cannot be observed in the secondary
pulse and the off-pulse except for observation~4, where it is seen in both.

\begin{table*}
\caption{$\chi^2$ and F-test values for fits to spectra of the main, the
secondary, and the off-pulse without, with one, and with two cyclotron
absorption lines.}
\label{tab_hnof}
\begin{tabular}{lccccccc} \multicolumn{2}{c}{} &
\multicolumn{3}{c}{Observation~1} & \multicolumn{3}{c}{Observation~2}\\
\hline & Main- & Sec.- & Off- & Main- & Sec.- & Off- \\ 
number of lines & DOF & Pulse& Pulse& Pulse& Pulse& Pulse& Pulse\\ 
\hline 
0 &(98 DOF)& 302.3 & 132.4 & 58.8 & 140.6 & 87.6 & 78.9\\ 
1 & (96 DOF)& 74.7 & 76.9 & 50.3 & 59.0 & --- & --- \\ 
$Q$(F) && 0.0\,\% & $<10^{-9}$\,\% & 0.06\,\%& 0.0\,\%&&\\ 
2 & (94 DOF)& 63.9 & --- & --- & --- & --- & ---\\ 
$Q$(F)& & 0.06\,\% & & & &&\\
\multicolumn{8}{c}{}\\ 
\multicolumn{8}{c}{}\\ 
\multicolumn{2}{c}{}& \multicolumn{3}{c}{Observation~3}&
\multicolumn{3}{c}{Observation~4}\\ 
\hline 
&& Main- & Sec.- & Off-& Main- & Sec.- & Off- \\ 
Number of lines &DOF& Pulse& Pulse& Pulse& Pulse& Pulse& Pulse \\ 
\hline 
0 &(98 DOF)& 131.2 & 68.2 & 93.1 & 339.7& 152.2 & 100.2 \\ 
1 &(96 DOF)& 79.8 & 53.2 & 63.5 & 87.6 &73.7 & 62.5 \\ 
$Q$(F)&& $10^{-8}$\,\%&0.0003\,\% & $<10^{-5}$\,\%& 0.0\,\% & $10^{-8}$\,\%
& $10^{-8}$\,\% \\ 
2&(94 DOF)& --- & --- & --- & 78.7 & 60.5& 57.9 \\ 
$Q$(F)&& & && 0.065\,\% &0.1\,\% &2.8\,\% \\
\end{tabular}
\end{table*}

\subsubsection{Pulse minus off-pulse}
The results from Sect.~\ref{sec_main} indicate that the cyclotron
absorption lines originate primarily in the pulsed emission and not in the
persistent flux. To further study this effect, we created a ``pulsed''
spectrum by accumulating the main or secondary pulse spectra and
subtracting the off-pulse spectra.  The results of fits to these spectra
are shown in Table \ref{tab_pmo}: the fundamental as well as the second
harmonic line (where observable) are deepest on the main pulse and less
deep, but still significant, in the secondary pulse.

The energy of the cyclotron absorption lines do not vary significantly
between observations. From the pulsed spectra we derive average line
energies of 24.2\,keV for the main pulse and 24.0\,keV for the secondary
pulse, the difference being well within the uncertainty.  This seems to be
even more the case as there does not seem to be any correlation between the
pulse phase and the line energy in the ten phase bins (see Sect.~\ref{ten}
and Fig.~\ref{fig_resvar}).

The coupling factor between the fundamental and second harmonic line is
greater than~2 in all four observations for those pulse phases where a
second line is observed.  We derive an average coupling factor of 2.4 which
agrees with the values previously obtained from the phase averaged spectra
and the main pulse spectra.

\begin{table}
\caption{Parameters of the cyclotron absorption lines in the ``pulsed''
spectra for all four observations.}
\label{tab_pmo}
\small\begin{tabular}{lr@{}lr@{}lr@{}lr@{}l}
\multicolumn{1}{c}{ } & \multicolumn{8}{c}{Main Pulse} \\ 
\hline 
Param. & \multicolumn{2}{c}{Obs.~1} & \multicolumn{2}{c}{Obs.~2}&
\multicolumn{2}{c}{Obs.~3}& \multicolumn{2}{c}{Obs.~4}\\ 
\hline 
$E_{\rm cyc}$ & 24.4&\err{0.4}{0.3} & 24.6&\err{1.2}{0.8} & 23.8&\err{1.4}{1.3}
&
24.0&\err{0.7}{0.6}\\ 
Depth 1 & 0.69&\err{0.06}{0.04} & 1.04&\err{0.57}{0.32} &
0.35&\err{0.12}{0.12} & 0.65&\err{0.12}{0.10}\\ 
Depth 2 & 2.21&\err{1.07}{1.63} & \multicolumn{2}{c}{---} & 0.57
&\err{0.83}{0.57}&
1.63&\err{1.33}{1.20}\\  
Factor & 2.55&\err{\infty}{0.12} & \multicolumn{2}{c}{---} &
2.15&\err{\infty}{\infty} &
2.53 & \err{0.69}{0.28}\\ 
\multicolumn{9}{c}{} \\ 
\multicolumn{1}{c}{ } & \multicolumn{8}{c}{Secondary Pulse} \\ 
\hline 
Param. & \multicolumn{2}{c}{Obs.~1}& \multicolumn{2}{c}{Obs.~2}&
\multicolumn{2}{c}{Obs.~3}& \multicolumn{2}{c}{Obs.~4} \\ 
\hline 
$E_{\rm cyc}$ & 23.3&\err{0.9}{0.9} & \multicolumn{2}{c}{---} &
\multicolumn{2}{c}{---} & 24.7&\err{1.3}{1.1} \\
Depth 1 & 0.17&\err{0.08}{0.09} & \multicolumn{2}{c}{---} &
\multicolumn{2}{c}{---} &
0.13&\err{0.04}{0.04} \\ 
Depth 2 & 1.79&\err{0.65}{0.64}& \multicolumn{2}{c}{---} &
\multicolumn{2}{c}{---} &
0.94&\err{0.35}{0.35} \\
Factor & 2.49&\err{0.19}{0.13} & \multicolumn{2}{c}{---} &
\multicolumn{2}{c}{---} &
2.10&\err{0.15}{0.17} \\  
\end{tabular}
\end{table}

\subsection{Comparison with other instruments}
\subsubsection{\textsl{Ginga}}
\label{ginga}
The \textsl{Ginga}-team used the \textsl{NPEX} model modified by
photoelectric absorption and an additive iron line at 6.4\,keV
(\cite{mihara95a}). They found a cyclotron absorption line at 24\,keV and
its second harmonic line. They used a fixed coupling constant of 2.0.

In order to compare our results with the \textsl{Ginga} observations, we
performed fits using the \textsl{NPEX} model.  Table~\ref{tab_npex} shows
the values found by \citey{mihara95a} using \textsl{Ginga} data and fits
with the \textsl{NPEX} model to our data. Finally, we compared this fit to
fits with the \textsl{FDCO} model to see which model describes the data
better.

Table~\ref{tab_npex} shows that our results are in very close agreement with
the \textsl{Ginga} observation of the fundamental cyclotron absorption
line. Both the \textsl{FDCO} and \textsl{NPEX} models provide very similar
line parameters. The only significant difference is that we find a coupling
factor of about 2.3 (independent of the model), while Mihara used the
canonical fixed value of 2.0.

\begin{table}
\caption{Comparison of the \textsl{NPEX} and Lorentzian cyclotron line
  parameters for phase average spectra found by \protect\citey{mihara95a}
  with \textsl{Ginga}, \protect\citey{orlandini98} with \textsl{BeppoSAX}
  and our observation~4 \textsl{RXTE} data.  We also show a fit to our data
  with the \textsl{FDCO} model. For the \textsl{Ginga} and \textsl{RXTE}
  data, the width of the second harmonic is related to that of the 
  fundamental by ``Factor.'' }
\label{tab_npex}
\begin{tabular}{lr@{}lr@{}lr@{}lr@{}l} 
Param. &\multicolumn{2}{c}{\textsl{Ginga}} & \multicolumn{2}{c}{{\em BeppoSAX}}
& \multicolumn{4}{c}{\textsl{RXTE}} \\
\cline{6-9}
& & & & & \multicolumn{2}{c}{\textsl{NPEX}} &\multicolumn{2}{c}{\textsl{FDCO}}
\\ 
\hline
$\alpha_1$ &0.61&\err{0.05}{0.05}&0.34&\err{0.11}{0.11}&0.42&\err{0.02}{0.02} &
0.94 &\err{0.04}{0.04}\\ 
$\alpha_2$ &$-2$&\ fixed&$-2.1$&\err{0.5}{0.5}&$-2.00$&\err{0.04}{0.08}&
\multicolumn{2}{c}{---}\\ 
$E_{\rm fold}$&6.4& \err{0.1}{0.1} &9.6&\err{1.8}{1.8}&7.3&\err{0.04}{0.08}&
\multicolumn{2}{c}{---}\\ 
E$_{\rm cyc}$ &24.5& \err{0.5}{0.5} &53&\err{2.0}{1.0}&23.4&\err{0.4}{0.3}&
23.3&\err{0.4}{0.3}\\ 
Depth 1&0.065&\err{0.015}{0.015} &1.5&\err{0.7}{0.6}&0.17&\err{0.02}{0.02}&
0.14&\err{0.02}{0.02}\\ 
Width 1 & 2.2 & \err{1.0}{1.0} & 20 & \err{4}{7} & 5 &\ fixed & 5&\ fixed \\
Depth 2 &0.80&\err{0.26}{0.26}&\multicolumn{2}{c}{---}&1.00&\err{0.08}{0.07}&
1.37&\err{0.28}{0.24}\\
Factor &2&\ fixed&\multicolumn{2}{c}{---}&2.33&\err{0.09}{0.08}&
2.38&\err{0.10}{0.10}\\ 
\hline 
\end{tabular}
\end{table}

\subsubsection{\textsl{HEXE} and \textsl{TTM} onboard {\sl Mir}}
By using both instruments in their work, \citey{kretschmar97c} were able to
analyze the broad energy range from 2 to 200\,keV, i.e., very similar to
\textsl{RXTE} but lacking the effective area especially in the lower energy
band. They used the standard model: a power law with high energy cutoff,
photoelectric absorption and an iron line. They found a cyclotron
absorption line at 22.6\,keV very similar to our value. They also find a
deep second harmonic absorption line which is coupled to the fundamental
line by a the fixed factor of 2.0. Fits to the \textsl{HEXE} and
\textsl{TTM} data using the \textsl{FDCO} model (Kretschmar 1997, priv.
comm.)  resulted in photon indices, cutoff energies, and folding energies
which are very similar to ours.

\subsubsection{\textsl{BeppoSAX}} 
\citey{orlandini98} analyzed Vela~X-1 spectra obtained with
\textsl{BeppoSAX}.  They simultaneously fit multi-instrument, broad-band
(2--100\,keV), phase-averaged spectra using the \textsl{NPEX} and
Lorentzian cyclotron absorption models. The resulting parameters are given
in Table~\ref{tab_npex}. While the \textsl{BeppoSAX} \textsl{NPEX}
parameters agree well with both \textsl{Ginga} and \textsl{RXTE}, no line
near 25\,keV was required to achieve an acceptable fit.  Based on this fact
and based on the \textsl{BeppoSAX} Phoswich Detector System (PSD) data from
20 to 100\,keV alone, they argued that the line at $\sim$54\,keV is the
fundamental one.  However, they report that allowing 2 absorption lines in
their fit, the first constrained to lie between 10 and 40\,keV, results in
a weak feature  at $\sim$24\,keV.  While this result was not
statistically significant, it makes clear that their data can admit a
feature near 25\,keV.  Further, their broad-band fits are based on
phase-averaged data only, while the present work and phase-resolved
\textsl{Ginga} (\cite{mihara95a}) analyses show that the line is strongest
during the pulse peaks. According to the results in Table~\ref{tab_npex},
the $\sim$25\,keV line is quite weak in the phase average, having an
optical depth only near 0.1.  Given that \textsl{HEXE}/\textsl{TTM},
\textsl{Ginga}, and now \textsl{RXTE} all detect a line near 25\,keV, it
seems most likely that this is the fundamental energy, corresponding to a
surface field of about $\rm 2\,10^{12}$\,G.  As \citey{orlandini98} note,
it is also possible that the line strength is variable with time, making
the lack of a detection with \textsl{BeppoSAX} less unlikely.

\section{Discussion and Summary}\label{discussion}
The \textsl{RXTE} observations of Vela~X-1 demonstrate the variability of
this source on many time scales. Besides the well known strong pulse to
pulse variations and the varying intensity states we observed a flare and
an interval with no observable X-ray pulsations.

In contrast, the fundamental accretion and emission geometry is obviously
very stable, as the comparison of the averaged pulse profiles within our
observations and with other results
(\cite{white83a,nagase83a,choi96a,orlandini98}) demonstrates.  Our energy
resolved pulse profiles show the transition from the complex five peak
structure at lower energies to the double peak above 15\,keV in
unprecedented detail. The determined pulse period shows a continuation of
the current overall spin-down trend.

One of the main goals of our observations -- to confirm or disprove the
existence of cyclotron absorption line features around $\sim$25\,keV and
$\sim$50\,keV has turned out to be more difficult than anticipated.  This
is due to the fact that the choice of the continuum model spectrum affects
the results significantly. As Sect.~\ref{models} demonstrates, the unwary
use of the ``classical'' high energy cutoff (\cite{white83a}) with its
abrupt onset of the cutoff can introduce line-like residuals and thus lead
to the introduction of artificial features in the modeling.  This fact,
which we have reported previously (\cite{kretschmar97b,kretschmar96b}), has
led us to study models with a smooth turnover, especially the \textsl{FDCO}
and \textsl{NPEX} models (Eqs.~\ref{eq_fdc} and~\ref{eq_npex}).  Both fit
the data equally well, so we have concentrated on the \textsl{FDCO} model
which has one parameter less and somewhat less freedom in its overall
shape.

Still -- regardless of the continuum description used -- we could only
achieve a convincing fit to the phase averaged spectra by including a
fundamental cyclotron line at 22--24\,keV into our models. According to the
F-test (Table~\ref{tab_avgftest}) this feature is significant for all
observations.  We therefore conclude that the fundamental line energy is
20--25\,keV in contrast to the results of \citey{orlandini98}. For two of
the four observations a harmonic feature is also observed, but with lesser
significance.  Contrary to the usual assumption, the second harmonic line
seems not be coupled by a factor of 2.0 in centroid energy and width to the
fundamental, but by a factor of 2.3 to 2.5 instead.
Note, however, that we can not formally rule out a factor of 2.0 and a
weak second harmonic due to limited fit statistics at higher energies
(cf.\ Sect.~\ref{pha}).

Confirming earlier results (\cite{kretschmar97c,mihara95a}), the pulse
phase resolved spectra indicate that, as with Her~X-1 (\cite{soong90a}),
the line features are deepest and most significant in the main pulse and
barely detectable outside the pulses.  Again, the best fit results are
obtained with a coupling factor of $>$2 between fundamental and second
harmonic.

At the moment we have no firm explanation for this surprising result.
Relativistic corrections will cause the resonance energies of the lines to
decrease, with increasing polar angle of the incident photons relative to
the magnetic field (\cite{harding91b}). The effect is more pronounced for
increasing harmonic number and thus should lead to a coupling factor $\le
2$. The possibility that we see contributions from two different regions at
different angles seems to be ruled out by the fact that we observe the
unusual coupling factor also in phase resolved spectroscopy.

If the emission region is significantly extended in height -- the
``fan-beam'' scenario -- the main contribution to the lines could in
principle be at different heights and therefore at different effective field
strengths (cf. \cite{burnard91a}). But it is not clear how this can be
reconciled with the simple double pulse structure and therefore probably
relatively simple emission geometry at these energies.

Another possibility, which needs further exploration, is that our results
are distorted by assuming too simple continuum models and line shapes. 
Monte-Carlo calculations (\cite{araya96a}) demonstrate that the
expected shapes of cyclotron lines deviate strongly from the simple
Lorentzian shape used throughout in spectral analysis. But, unfortunately,
currently no more complex model exists in suitable form for spectral 
analysis.

Although we are able to obtain acceptable fits for the data presented here,
there is an obvious need for better, self consistent models to fully
exploit the high quality data current instruments deliver.

\begin{acknowledgements}
  This research was supported by the Deutsche Agentur f\"ur
  Raumfahrtangelegenheiten (DARA) grants 50\,OR\,9205, 50\,OO\,9605,
  and 50\,OG\,9601, and by NASA grant NAG5-3272.
\end{acknowledgements}

\end{document}